\documentclass[runningheads]{llncs}
\usepackage{multirow}
\usepackage{hyperref}
\usepackage{subcaption}
\usepackage{graphicx}
\begin{document}
\title{On the Persistence of Persistent Identifiers \\ of the Scholarly Web}
%
%\titlerunning{The Memento Tracer Framework}
% If the paper title is too long for the running head, you can set an abbreviated paper title here
%
\author{
Martin Klein\inst{1,2} \and
Lyudmila Balakireva\inst{1,3}}
\authorrunning{M. Klein and L. Balakireva}
\institute{Los Alamos National Laboratory, Los Alamos, NM 87545, USA \and 
\email{mklein@lanl.gov} \\
\url{https://orcid.org/0000-0003-0130-2097} \and 
\email{ludab@lanl.gov} \\
\url{https://orcid.org/0000-0002-3919-3634} 
}
\maketitle              % typeset the header of the contribution
\begin{abstract}
Scholarly resources, just like any other resources on the web, are subject to reference rot 
as they frequently disappear or significantly change over time. Digital Object Identifiers 
(DOIs) are commonplace to persistently identify scholarly resources and have become 
the de facto standard for citing them. 
We investigate the notion of persistence of DOIs by analyzing their resolution on the web. 
We derive confidence in the persistence of these identifiers in part from the assumption 
that dereferencing a DOI will consistently return the same response, regardless of 
which HTTP request method we use or from which network environment we send the requests. 
Our experiments show, however, that persistence, according to our interpretation, is not 
warranted. We find that scholarly content providers respond differently to varying request 
methods and network environments and even change their response to requests against the same 
DOI. In this paper we present the results of our quantitative analysis that is aimed at 
informing the scholarly communication community about this disconcerting lack of consistency.
\end{abstract}
\section{Introduction}
The web is a very dynamic medium where resources frequently are being created, deleted, and 
moved \cite{bar-yossef:sic-transit,cho:evolution,cho:estimating}. Scholars have realized that, 
due to this dynamic nature, reliably linking and citing scholarly web resources is not a 
trivial matter \cite{lawrence:persistence,mccown:web_references}. 
Persistent identifiers such as the Digital Object Identifier (DOI)\footnote{\url{https://www.doi.org/}} 
have been introduced to address this issue and have become the de facto standard to persistently identify 
scholarly resources on the web. The concept behind a DOI is that while the location of a resource on the 
web may change over time, its identifying DOI remains unchanged and, when dereferenced on the web, 
continues to resolve to the resource's current location. This concept is based on the underlying assumption 
that the resource's publisher updates the mapping between the DOI and the resource's location if and when
the location has changed. If this mapping is reliably maintained, DOIs indeed provide a more persistent 
way of linking and citing web resources. 

While this system is not perfect \cite{bilder:doi_fail} and we have previously shown that authors of
scholarly articles often do not utilize DOIs where they should \cite{sompel:citation}, DOIs have become 
an integral part of the scholarly communication 
landscape\footnote{\url{https://data.crossref.org/reports/statusReport.html}}.
Our work is motivated by questions related to the consistency of resolving DOIs to scholarly content. 
From past experience crawling the scholarly web, for example 
in \cite{jones:content_drift,klein:one_in_five}, we have noticed that publishers do not necessarily 
respond consistently to simple HTTP requests against DOIs. We have instead observed 
scenarios where their response changes depending on what HTTP client and method is used. If we can 
demonstrate at scale that this behavior is common place in the scholarly communication landscape, it 
would raise significant concerns about the persistence of such identifiers for the scholarly web. In 
other words, we are driven by the question that if we can not trust that requests against the same DOI 
return the same result, how can we trust in the identifier's persistence?
In our previous study \cite{klein:who_is_asking} we reported the outcome of our initial investigation
into the notion of persistence of DOIs from the perspective of their behavior on the web. We found 
early indicators for scholarly publishers responding differently to different kinds of HTTP requests 
against the same DOI. 
In this paper we expand on our previous work by:
\begin{itemize}
\item re-executing the previous experiments with an improved technical setup,
\item adding additional experiments from a different network environment, 
\item adding additional experiments with different access levels to scholarly content, and
\item adding a comparison corpus to help interpret our findings and put them into perspective.
\end{itemize}
Adding these dimensions to our previous work and applying various different yet simple HTTP request 
methods with different clients to a large and arguably representative corpus of DOIs, we address
the following research questions:
\begin{enumerate}
\item What differences in dereferencing DOIs can we detect and highlight?
\item In what way (if at all) do scholarly content providers' responses change depending on network
environments?
\item How do observed inconsistencies compare to responses by web servers providing popular 
(non-scholarly) web content?
\item What effect do Open Access and non Open Access content providers have on the overall picture?
\item What is the effect of subscription levels to the observed inconsistencies?
\end{enumerate}
These five research questions (RQs) aim at a quantitative analysis of the consistency of HTTP responses.
We do not claim that such consistency is the only factor that contributes to persistence of scholarly
resource identifiers. We argue, however, that without a reassuring level of consistency, our trust 
in the persistence of an identifier and its resolution to a resource's current location is significantly 
diminished. 

In the remainder of this paper we will briefly highlight previous related work 
(Section \ref{sec:related_work}), outline the experiments' setup 
(Section \ref{sec:experimental_setup}), and address our research questions 
(Section \ref{sec:experimental_results}) before drawing our conclusions
(Section \ref{sec:conclusions}).
\section{Related Work} \label{sec:related_work}
DOIs are the de facto standard for identifying scholarly resources on the web, supported by traditional
scholarly publishers as well as repository platforms such as Figshare and Zenodo, for example.
When crawling the scholarly web for the purpose of aggregation, analysis, or archiving, DOIs are therefore 
often the starting point to access resources of interest. The use of DOIs for references in scholarly 
articles, however, is not as wide-spread as it should be. In previous work \cite{sompel:citation}, 
we have presented evidence that authors often use the URL of a resource's landing page rather than its
DOI when citing the resource. This situation is undesirable as it requires unnecessary deduplication 
for efforts such as metrics analysis or crawling.
These findings were confirmed in a large scale study by Thompson and Jian \cite{thompson:common_crawl}
based on two samples of the web taken from Common Crawl\footnote{\url{http://commoncrawl.org/}} datasets.
The authors were motivated to quantify the use of HTTP DOIs versus URLs of landing pages in these two
samples generated from two snapshots in time. They found more than 5 million actionable HTTP DOIs in 
the first dataset from 2014 and about $10\%$ of them in the second dataset from 2017 but identified as 
the corresponding landing page URL, not the DOI.
It is worth noting that not all resources referenced in scholarly articles have a DOI assigned to them
and are therefore subject to typical link rot scenarios on the web. In large-scale studies, we have
previously investigated and quantified the ``reference rot'' phenomenon in scholarly 
communication \cite{jones:content_drift,klein:one_in_five} focusing on ``web at large'' resources that
do not have an identifying DOI. 

Any large-scale analysis of the persistence of scholarly resources requires machine access as human
evaluations typically do not scale. 
Hence, making web servers that serve (scholarly) content more friendly to machines has been the focus of
previous efforts by the digital library community with the agreement that providing accurate and 
machine-readable metadata is a core requirement \cite{brandman:crawler_friendly,nelson:harvesting}.
To support these efforts, recently standardized frameworks are designed to help machines synchronize 
metadata and content between scholarly platforms and repositories \cite{klein:technical_framework}.

The study by Alam et al. \cite{alam:methods} is related to ours in the way that the authors 
investigate the support of various HTTP request methods by web servers serving popular web pages.
The authors issue OPTIONS requests and analyze the values of the ``Allow'' response header to 
evaluate which HTTP methods are supported by a web server. The authors conclude that a sizable 
number of web servers inaccurately report supported HTTP request methods. 
\section{Experimental Setup} \label{sec:experimental_setup}
\subsection{Dataset Generation}
To the best of our knowledge, no dataset of DOIs that identify content representative of the diverse scholarly 
web is available to researchers. Part of the problem is the scale and diversity of the publishing industry
landscape but also the fact that the Science, Technology, and Medicine (STM) market is dominated by a few large 
publishers \cite{johnson:stmreport}.
We therefore reuse the dataset generated for our previous work \cite{klein:who_is_asking} that consists of 
$10,000$ randomly sampled DOIs from a set of more than $93$ million DOIs crawled by the Internet Archive. 
We refer to \cite{klein:who_is_asking} for a detailed description of the data gathering process, an analysis
of the composition of the dataset, and a discussion of why we consider this dataset to be representative of 
the scholarly landscape.
In addition, to be able to put our findings from the DOI-based dataset in perspective, we created a dataset of 
the top $10,000$ most popular URIs on the web as extracted from the freely available ``Majestic Million'' 
index\footnote{\url{https://blog.majestic.com/development/majestic-million-csv-daily/}} on November 14, 2019. 
%We refer to this dataset as $\textbf{Web}$. 
%
%
\subsection{HTTP Requests, Clients, and Environments}
HTTP transactions on the web consists of a client request and a server response. As detailed in RFC 
7231 \cite{http:rfc7231} requests contain a request method and request headers and responses contain 
corresponding response headers. GET and HEAD are two of the most common HTTP request methods (also detailed 
in RFC 7231). The main difference between the two methods is that upon receiving a client request with the 
HEAD method, a server only responds with its response headers but does not return a content body to the client. 
Upon receiving a client request with the GET method, on the other hand, a server responds by sending the 
representation of the resource in the response body in addition to the response headers.  

It is important to note that, according to RFC 7231, we should expect a server to send the same headers
in response to requests against the same resource, regardless whether the request is of type HEAD or GET. 
RFC 7231 states: ``The server SHOULD send the same header fields in response to a HEAD request as it would 
have sent if the request had been a GET...''.

To address our research questions outlined earlier, we utilize the same four methods described 
in \cite{klein:who_is_asking} to send HTTP requests: 
\begin{itemize}
\item \textbf{HEAD}, a HEAD request with cURL\footnote{A popular lightweight HTTP client for the command 
line interface \url{https://curl.haxx.se/}.} 
\item \textbf{GET}, a simple GET request with cURL
\item \textbf{GET+} a GET request that includes typical browsing parameters such as user agent and accepted 
cookies with cURL 
\item \textbf{Chrome}, a GET request with Chrome\footnote{Web browser controlled via the Selenium 
WebDriver \url{https://selenium.dev/projects/}.}
\end{itemize}

We sent these four requests against the HTTPS-actionable format of a DOI, meaning the form of 
\texttt{https://doi.org/<DOI>}. This is an important difference to our previous work 
(\cite{klein:who_is_asking}) where we did not adhere to the format recommended by the DOI
Handbook\footnote{\url{https://www.doi.org/doi_handbook/3_Resolution.html}}.
For the first set of experiments and to address RQ1, we send these four HTTP requests against each of the 
$10,000$ DOIs from an Amazon Web Services (AWS) virtual machine located at the U.S. East Coast. The 
clients sending the requests are therefore not affiliated with our home institution's network. Going forward, 
we refer to this external setup as the $DOI_{ext}$ corpus.
In addressing RQ2, we anticipate possible discrepancies in HTTP responses from servers depending on the network 
from which the request is sent. Hence, for the second set of experiments, we send the same four requests to the 
same $10,000$ DOIs from a machine hosted within our institution's network. Given that the machine's IP address
falls into a range that conveys certain institutional subscription and licensing levels to scholarly publishers, 
this internal setup, which we refer to going forward as $DOI_{int}$, should help surface possible differences.
To address RQ3 we compare our findings to responses from servers providing non-scholarly content by sending 
the same four requests against each of the $10,000$ URIs from our dataset of popular websites. From here on,
we refer to this corpus as the $Web$ dataset. 
\section{Experimental Results} \label{sec:experimental_results}
In this section we report our observations when dereferencing HTTPS-actionable DOIs with our four methods.
Each method automatically follows HTTP redirects and records information about each link in the redirect chain.
For example, a HEAD request against \url{https://doi.org/10.1007/978-3-030-30760-8_15} results in a redirect
chain consisting of the following links:
\begin{enumerate}
\item \url{http://link.springer.com/10.1007/978-3-030-30760-8_15}
\item \url{https://link.springer.com/10.1007/978-3-030-30760-8_15}
\item \url{https://link.springer.com/chapter/10.1007%2F978-3-030-30760-8_15}
\end{enumerate}
with the last one showing the $200~OK$ response code. Note that only the first redirect comes from the server at 
doi.org (operated by the Corporation for National Research Initiatives 
(CNRI)\footnote{\url{https://www.cnri.reston.va.us/}}) and it points to the appropriate location on the publisher's 
end. All consecutive redirects remain in the same domain and, unlike the HTTP DOI, are controlled by the publisher.

It is important to note that all four methods are sent with the default timeout of $30$ seconds, meaning
the request times out if a server does not respond within this time frame. In addition, all methods are
configured to follow a maximum of $20$ redirects. 
\subsection{Final Response Codes} \label{sec:final_response_codes}
The first aspect of consistency, as projected onto our notion of persistence, we investigate is the response code 
of the last accessible link in the redirect chain when dereferencing DOIs (or URIs in the case of the $Web$ corpus). 
Intuitively and informed by our understanding of persistence, we expect DOIs as persistent identifiers return the 
same response code to all issued requests, regardless of the request method used. 

Table \ref{tab:final_codes} summarizes the response codes for our three different corpora and the four different 
methods for each of them. The frequency of response codes (in percent) is clustered into 200-, 300-, 400-, and 
500-level columns, plus an error column. The latter represents requests that timed out and did not return any 
response or response code.
The first main observation from Table \ref{tab:final_codes} is that the ratio of response codes for all four methods
and across all three corpora is inconsistent. Even within individual corpora, we notice significant differences.
For example, for the $DOI_{ext}$ corpus we see $40\%$ and $24\%$ of GET and GET+ requests respectively end in 
300-level response codes. We consider this number particularly high as the vast majority of these responses have a
$302~Found$ status code that indicates further action needs to be taken by the client to fulfill the request, 
for example, send a follow-up request against the URI provided in the Location header field 
(see RFC 7231 \cite{http:rfc7231}). In other words, no HTTP request (and redirect chain) should end with such 
a response code. 
A different reason for these observations could be a server responding with too many consecutive 300-level responses, 
causing the client to stop making follow-up requests (the default for our methods was $20$ requests). However, we 
only recorded this behavior a few times and it therefore can not explain these high numbers.
Another observation for the same corpus is the fairly high ratios for 400-level responses, 
particularly for HEAD requests. The fact that this number ($12.58\%$) is two to three times as high as for the 
other three requests for the same corpus is noteworthy. 
\begin{table}
\centering
\caption{Final HTTP response codes, aggregated into five levels, following the DOI/URI redirect chain}
\begin{tabular}{|c|c||c|c|c|c|c|} \hline
\textbf{Corpus} & \textbf{Method} & \textbf{2xx} & \textbf{3xx} & \textbf{4xx} & \textbf{5xx} & \textbf{Err}\\ \hline \hline
\multirow{4}{*}{$DOI_{ext}$} 
& HEAD & 75.4 & 9.93 & 12.58 & 2.09 & 0 \\
& GET & 53.07 & 40.49 & 6.06 & 0.06 & 0.32 \\
& GET+ & 70.71 & 24.34 & 4.58 & 0.05 & 0.32 \\
& Chrome & 87.79 & 6.17 & 5.94 & 0.1 & 0 \\ \hline \hline
\multirow{4}{*}{$DOI_{int}$} 
& HEAD & 70.64 & 16.98 & 8.85 & 3.52 & 0.01 \\
& GET & 76.13 & 16.66 & 5.71 & 1.48 & 0.02 \\
& GET+ & 80.29 & 15.26 & 4.04 & 0.41 & 0 \\
& Chrome & 90.2 & 5.95 & 3.57 & 0.18 & 0.1 \\ \hline \hline
\multirow{4}{*}{$Web$} 
& HEAD & 70.69 & 4.86 & 5.63 & 1.32 & 17.5 \\
& GET & 56.71 & 5.35 & 2.78 & 0.6 & 34.56 \\
& GET+ & 57.43 & 5.54 & 1.87 & 0.52 & 34.64 \\
& Chrome & 74.8 & 4.56 & 2.66 & 0.65 & 17.33 \\ \hline
\end{tabular}
\label{tab:final_codes}
\end{table}

Except for HEAD requests, the ratio of 300-level responses decreased for the $DOI_{int}$ corpus. We do see more 
$301~Moved~Permanently$ responses in this corpus compared to $DOI_{ext}$ but given that this fact should not have 
a different impact for individual request methods, we can only speculate why the ratio for HEAD requests went up.
The ratio of 400-level responses is not insignificant in both corpora and it is worth noting that this category is 
dominated by the $403$ response code, which means a server indicates to a client that access to the requested URI 
is forbidden. This response would make sense for requests to resources for which we do not have institutional 
subscription rights or licensing agreements, for example, but then we would expect to see these numbers being 
consistent for all methods.

As a comparison, the requests for the $Web$ corpus seem to mostly result in one of two columns. Either they 
return a 200-level response or an error (no response code at all). The ratios in the error category are particularly 
high for the GET and the GET+ methods at around $34\%$.
\subsection{Redirect Chain}
\begin{figure}[h!]
    \centering
    \begin{subfigure}[h]{0.95\textwidth}
    \centering
        \includegraphics[scale=0.4]{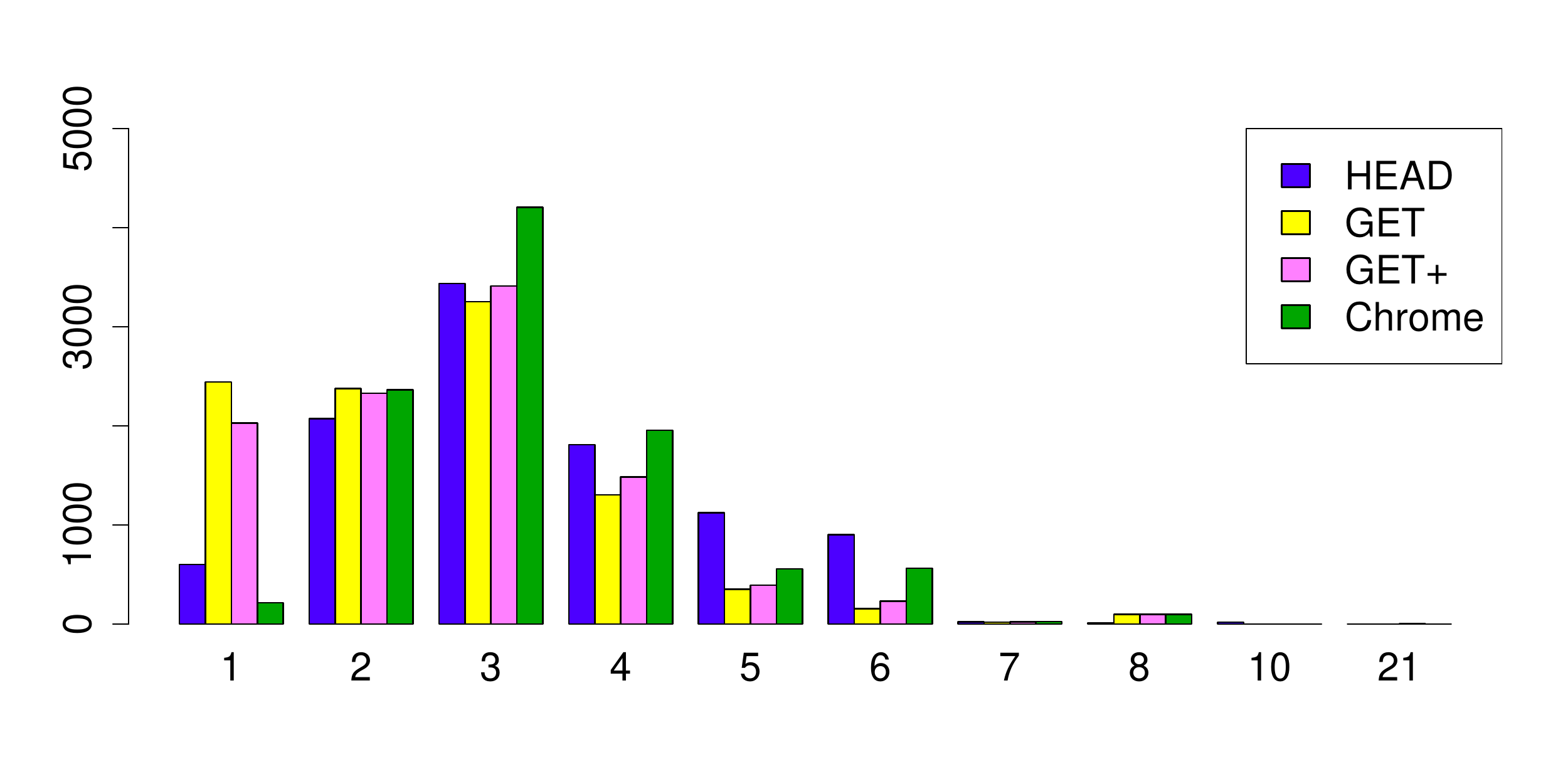}
        \caption{$DOI_{ext}$ corpus}
        \label{fig:redirs_ext}
    \end{subfigure}
    ~
    \begin{subfigure}[h]{0.95\textwidth}
    \centering
        \includegraphics[scale=0.4]{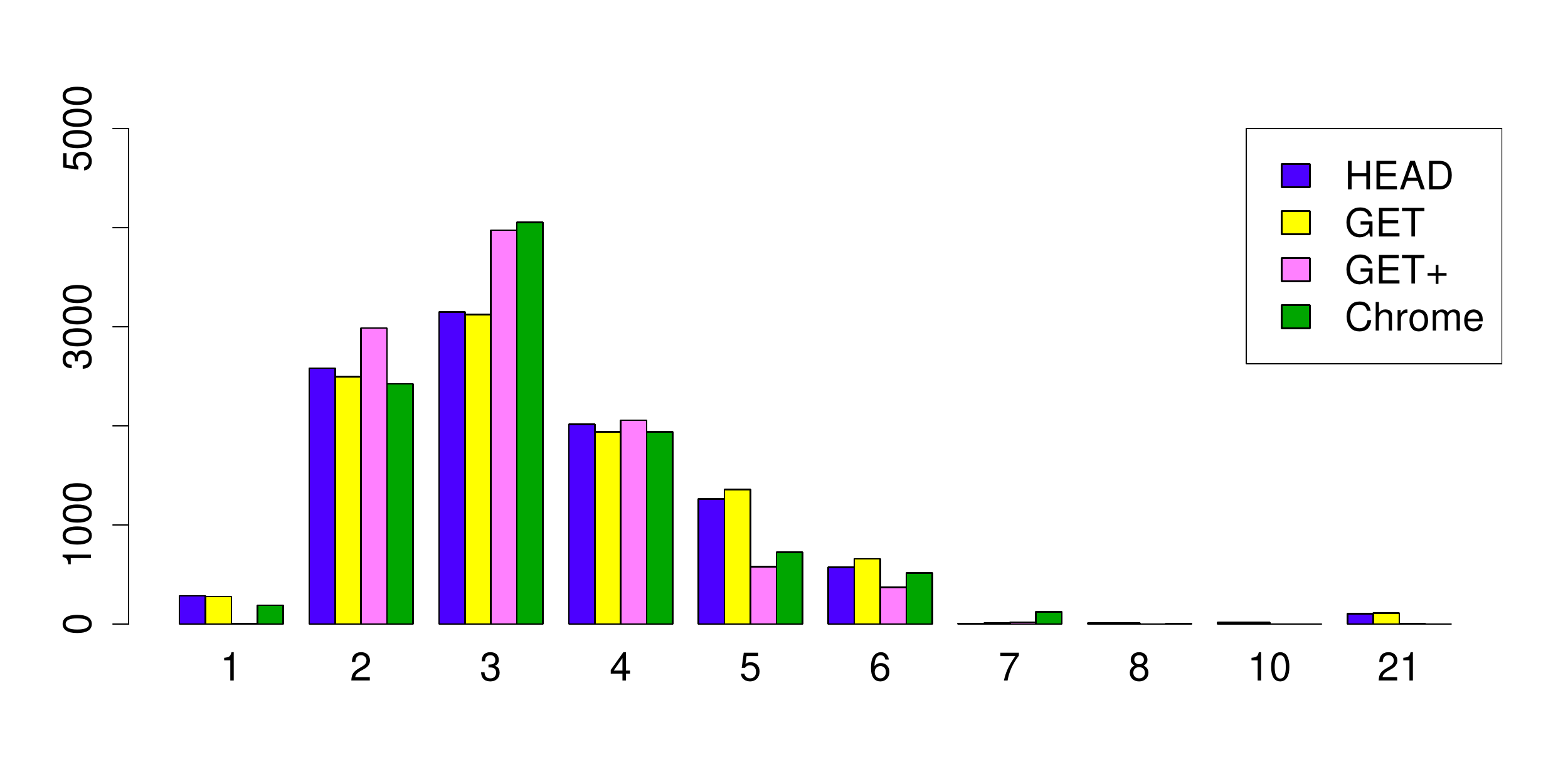}
        \caption{$DOI_{int}$ corpus}
        \label{fig:redirs_int}
    \end{subfigure}
    ~
    \begin{subfigure}[h]{0.95\textwidth}
    \centering
        \includegraphics[scale=0.4]{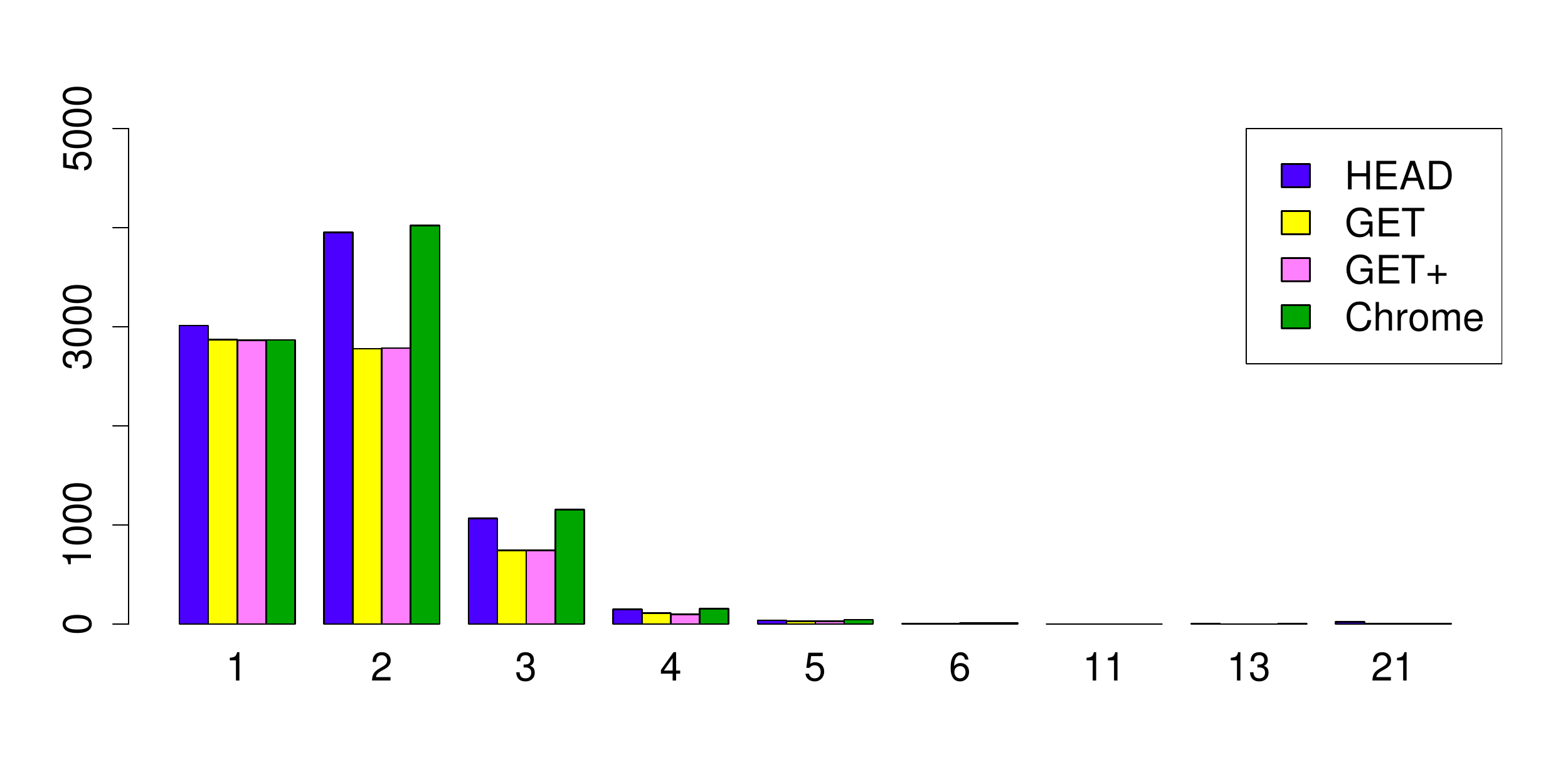}
        \caption{$Web$ corpus}
        \label{fig:redirs_web}
    \end{subfigure}
    \caption{Number of total links in DOI/URI redirect chains per corpus}
    \label{fig:redirs}
\end{figure}

The next aspect of persistence in our investigation is the overall length of the redirect chain when dereferencing 
DOIs. Intuitively speaking, we expect the chain length to be the same for persistent identifiers, regardless of 
the HTTP method used.  
Figure \ref{fig:redirs} shows histograms of chain lengths distinguished by corpora and request methods. Note that 
the reported lengths are independent of the final response code reported earlier and that DOIs/URIs that resulted 
in errors are excluded from this analysis. Figure \ref{fig:redirs_ext} shows the observed chain lengths for the 
$DOI_{ext}$ corpus. We note that the distribution of chain lengths is not equal among request methods. The GET and 
GET+ methods, for example, are much more strongly represented at length one than either of the other methods. 
Generally speaking however, lengths two, three, and four represent the majority for the requests in the $DOI_{ext}$ 
corpus.

The same holds true for the $DOI_{int}$ corpus (shown in Figure \ref{fig:redirs_int}) but we notice the frequency of
length one has almost disappeared. When comparing the two corpora, we observe that the Chrome method shows fairly 
consistent frequencies of redirect chain length and most often results in length three.

Figure \ref{fig:redirs_web} offers a comparison by showing the redirect chain lengths of dereferencing URIs from the 
$Web$ corpus. We see a significant shift to shorter redirect chains with the majority being of length one or two. 
While we recorded chains of length four and beyond, these occurrences were much less frequent. The HEAD and Chrome 
methods appear to be well-aligned for all observed lengths. 
It is worth mentioning that we recorded chain length beyond our set maximum of $20$ (indicated as $21$ in the figures).
We question the reasoning for such responses but leave a closer analysis of these extensive redirect chains for future 
work. 
\subsection{Changing Response Codes}
The third aspect of our investigation centers around the question whether HTTP response codes change,
depending on what HTTP request method is used. We have shown in Section \ref{sec:final_response_codes}
that dereferencing DOIs does not result in the same response codes but varies depending on what request method
we used. In this section we analyze the nature of response code change per DOI and request method.
This investigation aims at providing clarity about if and how response codes change and the ramifications 
for the notion of persistence.

\begin{figure}[t!]
    \centering
    \begin{subfigure}[h]{0.49\textwidth}
        \includegraphics[scale=0.25]{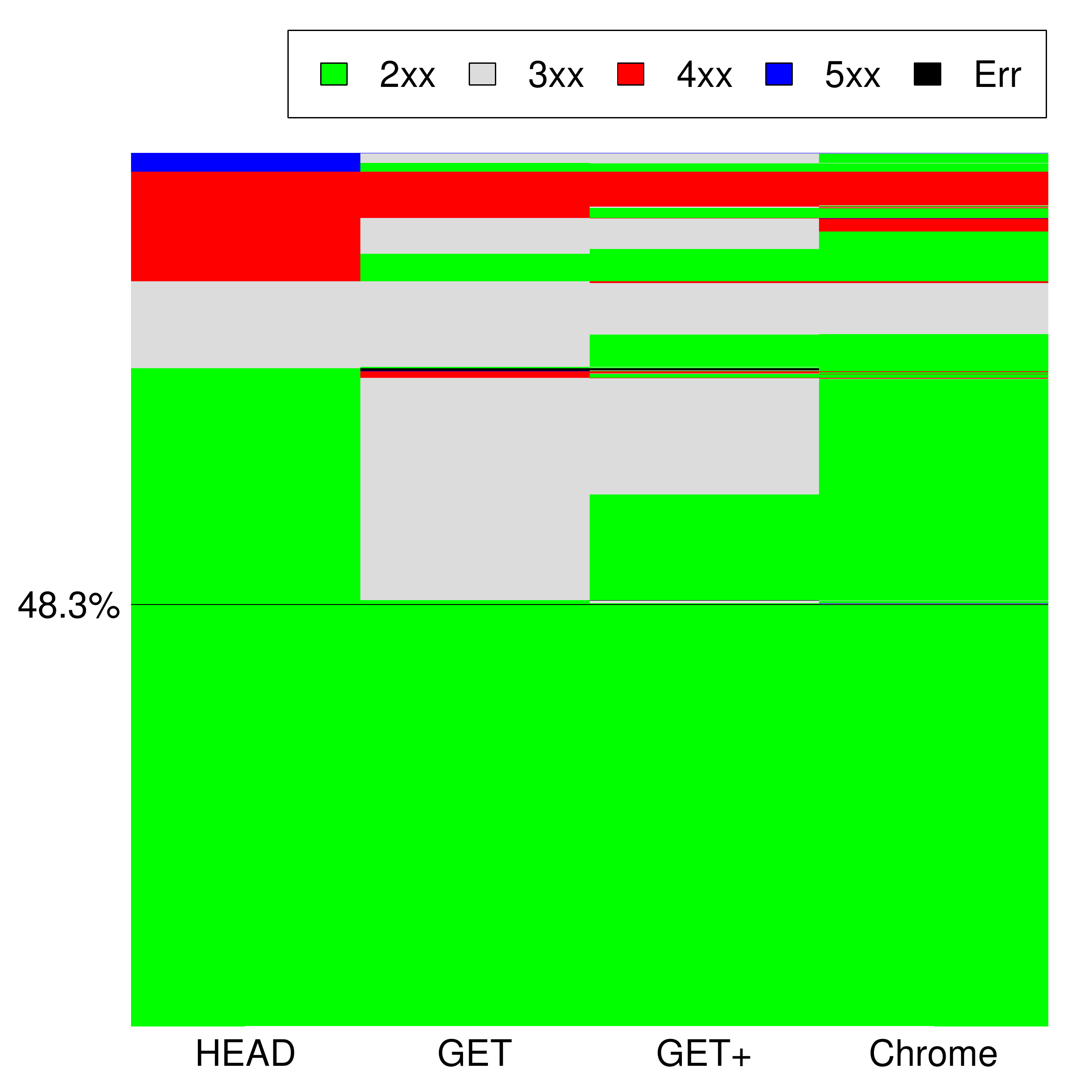}
        \caption{$DOI_{ext}$ corpus}
        \label{fig:final_codes_by_DOI_ext}
    \end{subfigure}
    %~
    \begin{subfigure}[h]{0.49\textwidth}
        \includegraphics[scale=0.25]{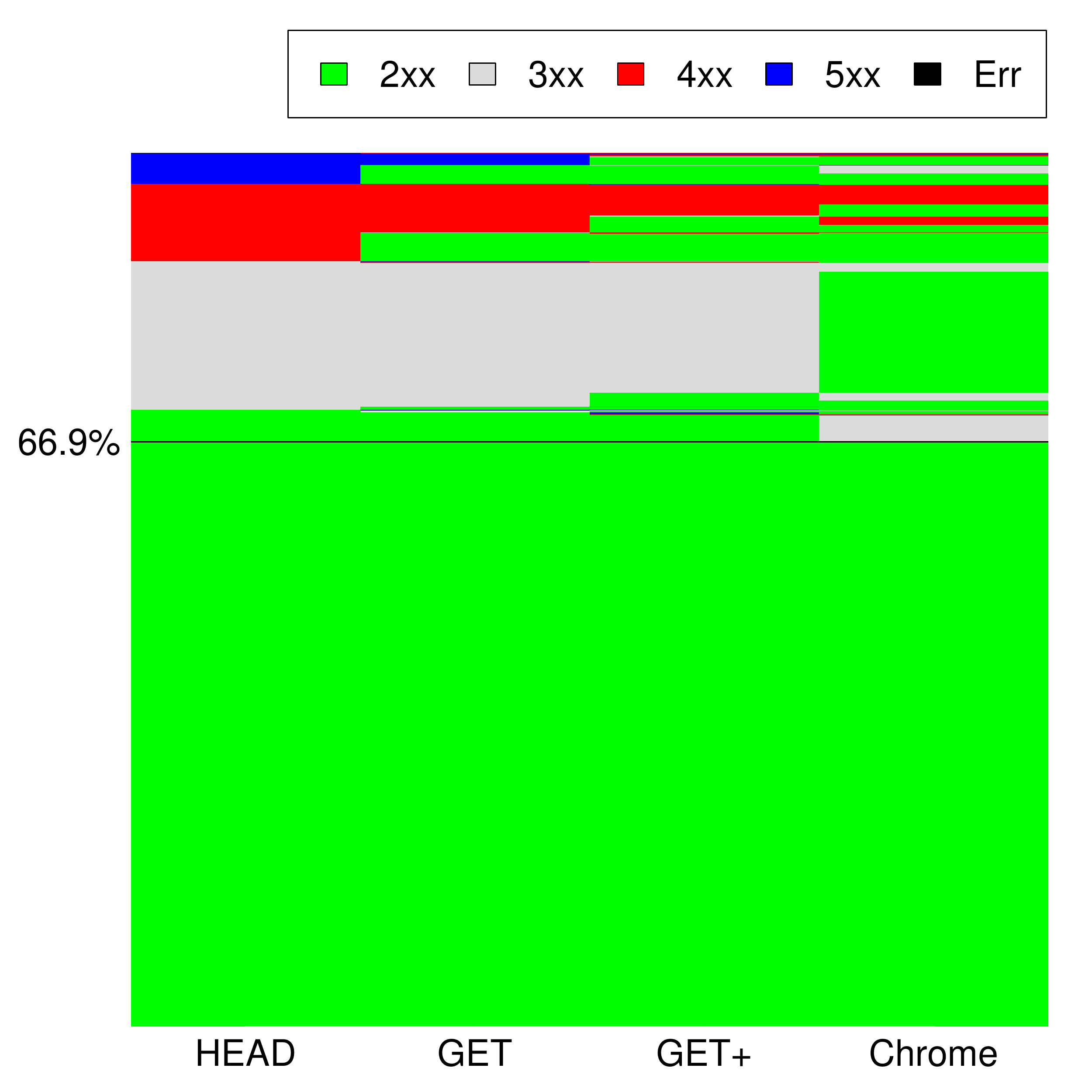}
        \caption{$DOI_{int}$ corpus}
        \label{fig:final_codes_by_DOI_int}
    \end{subfigure}
    ~
    \begin{subfigure}[h]{0.5\textwidth}
        \includegraphics[scale=0.25]{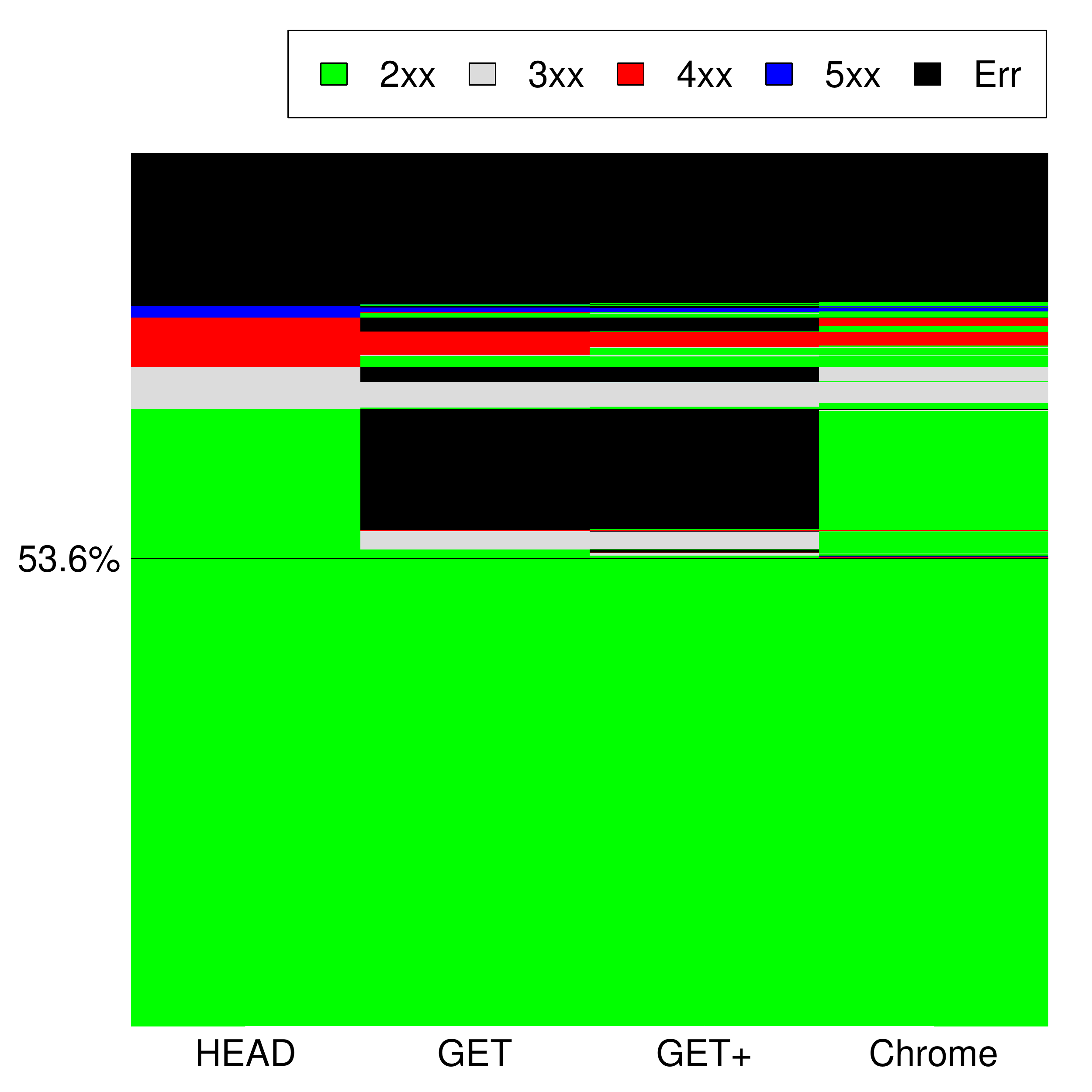}
        \caption{$Web$ corpus}
        \label{fig:final_codes_by_DOI_web}
    \end{subfigure}
    \caption{Final HTTP response codes by DOI/URI per corpus}
    \label{fig:final_codes_by_DOI}
\end{figure}

Figure \ref{fig:final_codes_by_DOI} shows all response codes again binned into 200- (green), 300- (gray), 400- (red), 
500-level (blue), and error (back) responses per DOI for all three corpora. The request methods are represented 
on the x-axis and each of the $10,000$ DOIs is displayed on the (unlabeled) y-axis.
Figure \ref{fig:final_codes_by_DOI_ext} shows the response codes and their changes from one method to another for the 
$DOI_{ext}$ corpus. 
We see that merely $48.3\%$ of all $10,000$ DOIs consistently return a 200-level response, regardless of which request
method is used. This number is surprisingly low. The fact that, consistently across request methods, more than half of 
our DOIs fail to successfully resolve to a target resource strongly indicates that the scholarly communication 
landscape is lacking the desired level of persistence. 
We further see major differences in response codes depending on the request method. For example, a large portion, 
just over $40\%$, of all DOIs return a 300-level response for the simple GET request. However, $12\%$ of these DOIs 
return a 200-level response with any of the other three request methods and $25\%$ return a 200-level response if 
only the HEAD or Chrome method is used. We further find $13\%$ of DOIs resulting in a 400-level response with the 
HEAD request but of these only $30\%$ return the same response for any of the other request methods. In fact, $25\%$ 
of them return a 200-level response when any other request method is used.
Without further analysis of the specific links in the redirect chain and their content, which we leave for future work, 
we can only hypothesize that web servers of scholarly content take the request method into consideration and respond 
accordingly when resolving DOIs. However, this lack of consistency is worrisome for everyone concerned about persistence 
of the scholarly record. 

Figure \ref{fig:final_codes_by_DOI_int} shows our findings from the $DOI_{int}$ corpus. We see the numbers improved, most
noticeably with $66.9\%$ of DOIs returning a 200-level response across the board. However, we still find almost $14\%$ of 
DOIs returning a 300-level response for the first three and a 200-level response only for our Chrome method. 
We also see a similar ratio of 400-level responses for the HEAD method that decreases with the GET, GET+, and Chrome methods,
similar to our observation for the $DOI_{ext}$ corpus.
The ratio of 500-level responses slightly increased from $2\%$ in the previous corpus to $3.5\%$ here. However, here too
the majority of those DOIs return a different response code when methods other than HEAD are used.
The observations from Figure \ref{fig:final_codes_by_DOI_int} show that even requests sent from within a research institution 
network are treated differently by scholarly content providers and, depending on the request method used, the level of
consistency suffers.

Figure \ref{fig:final_codes_by_DOI_web} shows the numbers for the $Web$ corpus and therefore offers a comparative picture 
to our above findings. For the $Web$ corpus we see $53.6\%$ of all $10,000$ URIs returning a 200-level response code, which
is ahead of the $DOI_{ext}$ but well below the $DOI_{int}$ corpus numbers. We further see $17\%$ of URIs returning an error,
regardless of the request. We can only speculate about the reasons for this high number of unsuccessful requests but our
best guess is that web servers of these popular websites have sophisticated methods in place that detect HTTP requests 
sent from machines and simply do not send a response when detected. This even holds true for our Chrome method, which closely
resembles a human browsing the web. 
Not unlike what we have seen in the $DOI_{ext}$ corpus the $Web$ corpus shows $15\%$ of requests not being successful 
with the GET and GET+ methods but being successful (200-level response) with the HEAD and Chrome methods. 
These findings indicate that popular but not necessarily scholarly content providers also send responses depending on the 
request method. However, we see fewer 300-, 400-, and 500-level responses for this corpus.
\subsection{Responses Depending on Access Level}
The distinction between the $DOI_{ext}$ and $DOI_{int}$ corpora serves to highlight patterns for the lack of consistent
responses by scholarly publishers when accessed from outside and within an institutional network. Our observations 
raise further questions about possible differences between access levels. In particular, we are motivated to evaluate 
the responses for:
\begin{itemize}
\item DOIs identifying Open Access ($OA$) content versus their non-OA counterparts ($nOA$) and
\item DOIs identifying content to which we have access due to institutional subscription and licensing agreements
($SUB$) versus those we do not ($nSUB$).
%($SUB$) versus those we do not have access to ($nSUB$).
\end{itemize}
We utilize our $DOI_{ext}$ corpus to analyze responses of DOIs identifying OA content and the $DOI_{int}$ corpus to 
investigate responses for DOIs that lead to licensed content. Identifying OA content can be a non-trivial task but rather 
than manually inspecting all of the $10,000$ DOIs, we rely on the popular unpaywall service and their 
API\footnote{\url{https://unpaywall.org/products/api}} to determine whether a DOI identifies OA content. To identify
licensed content, we match institutional subscription information to base URIs of dereferenced DOIs. 
Table \ref{tab:access_levels} summarizes the resulting numbers of DOIs and their access levels in our corpora. We
realize that the numbers for licensed content may not be representative as other institutions likely have different
subscription levels to scholarly publishers. However, given that we consider our DOI corpus representative, we are
confident the ratios represent a realistic scenario. 

\begin{figure}[t!]
    \centering
    \begin{subfigure}{0.49\textwidth}
        \includegraphics[scale=0.25]{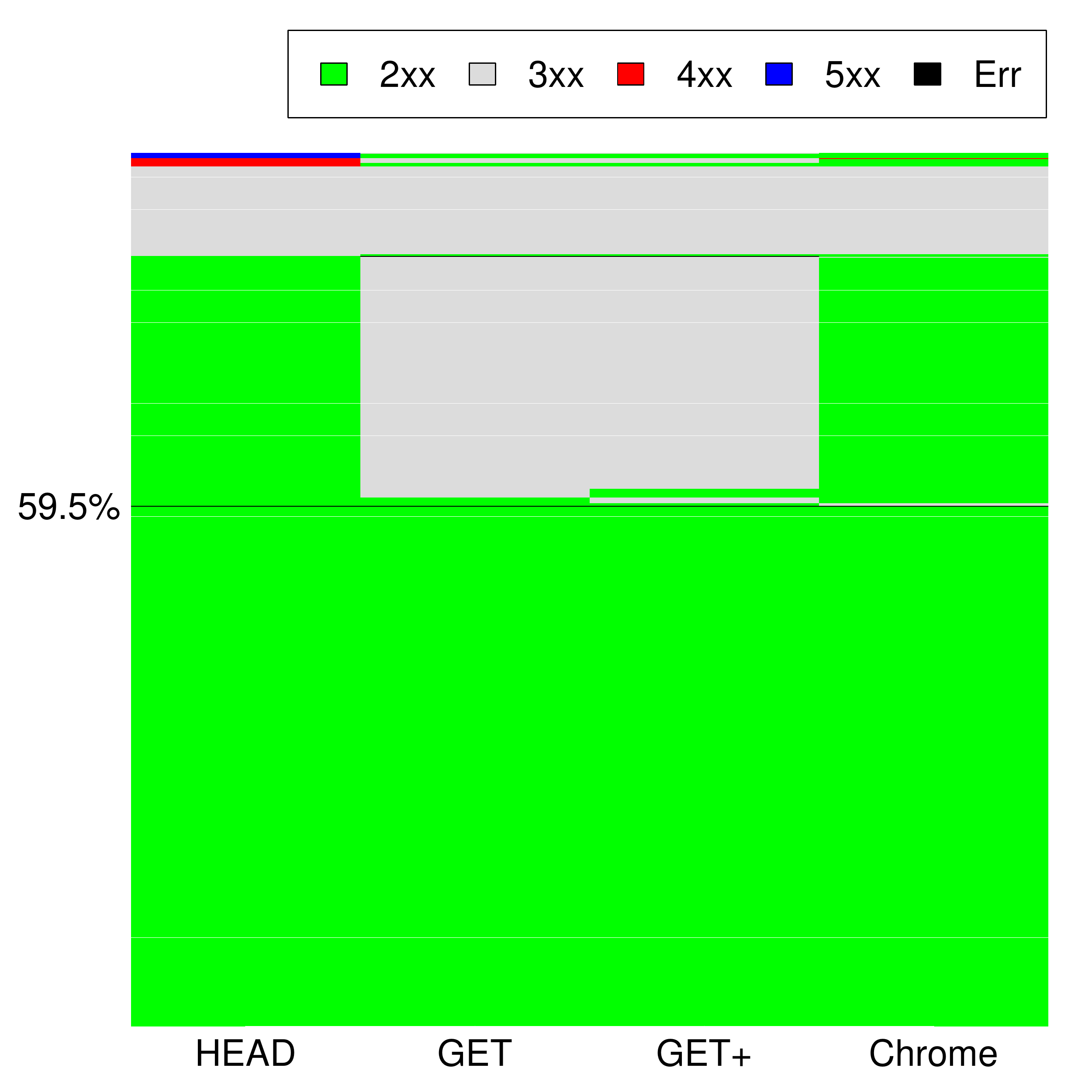}
        \caption{OA articles}
        \label{fig:final_codes_by_DOI_ext_OA}
    \end{subfigure}
    %~
    \begin{subfigure}{0.49\textwidth}
        \includegraphics[scale=0.25]{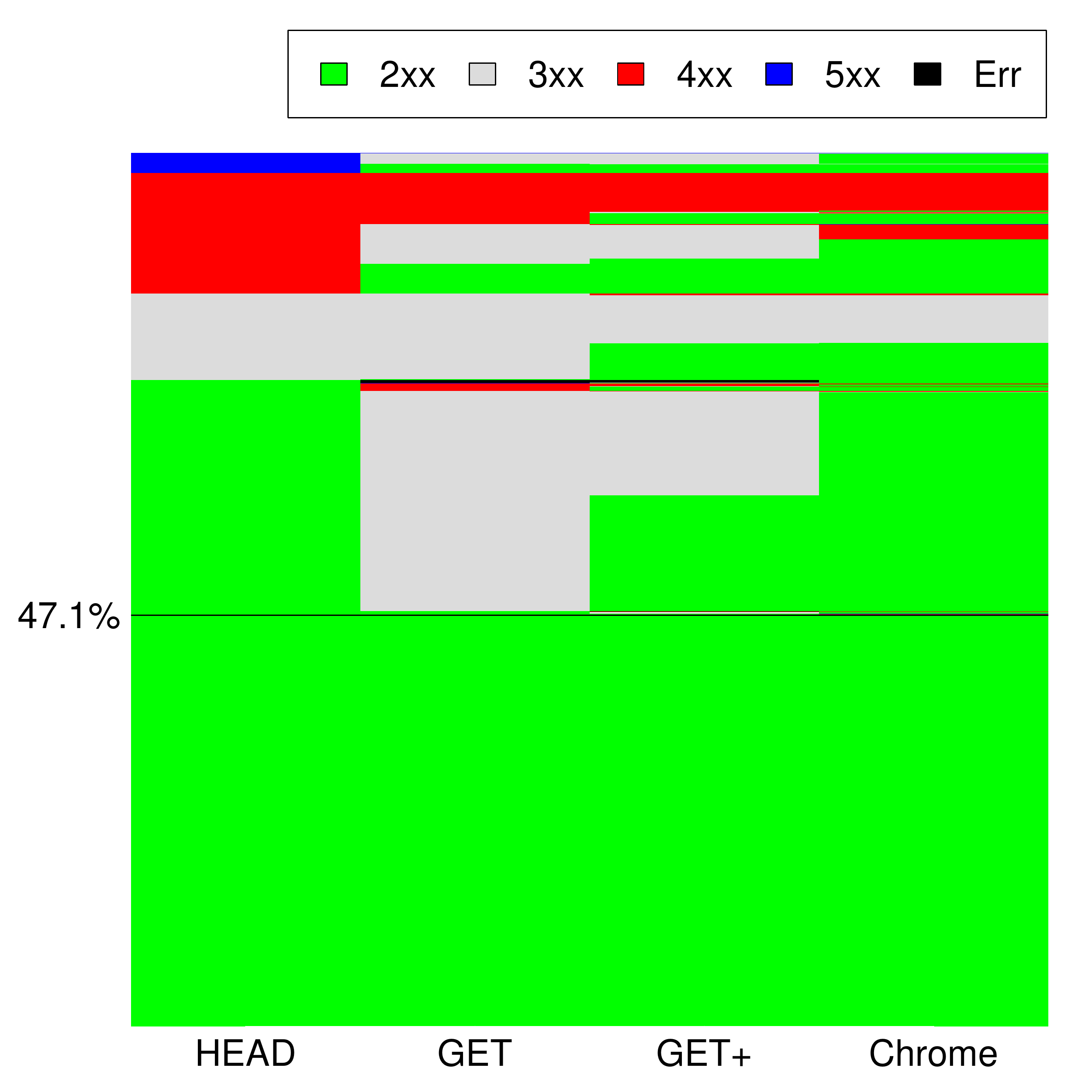}
        \caption{Non-OA articles}
        \label{fig:final_codes_by_DOI_ext_nOA}
    \end{subfigure}
    \caption{$DOI_{ext}$ final HTTP response codes distinguished by OA and nOA}
    \label{fig:final_codes_by_DOI_ext_OAnOA}
\end{figure}

Figure \ref{fig:final_codes_by_DOI_ext_OAnOA} shows the final response codes for the $DOI_{ext}$ corpus, similar in
style to Figure \ref{fig:final_codes_by_DOI}, with the DOIs along the y-axis and our four request methods
on the x-axis. Figure \ref{fig:final_codes_by_DOI_ext_OA} shows the response codes for the $973$ OA DOIs and 
Figure \ref{fig:final_codes_by_DOI_ext_nOA} shows the remaining $9.027$ DOIs that identify non-OA content. 
The first observation we can make from these two figures is that OA DOIs return 200-level responses for all
requests more often than non-OA DOIs with $59.5\%$ versus $47.1\%$. We can further see that even for OA DOIs
the GET and GET+ method do not work well. $26\%$ of DOIs return a 300-level response for these two methods but return
a 200-level response for the HEAD and Chrome methods. 
If we compare Figure \ref{fig:final_codes_by_DOI_ext_OAnOA} with \ref{fig:final_codes_by_DOI_ext} we can see a 
clear resemblance between Figure \ref{fig:final_codes_by_DOI_ext}, the figure for the overall corpus, and 
Figure \ref{fig:final_codes_by_DOI_ext_nOA}, the figure for non-OA DOIs. Given the fact that we have many more 
non-OA DOIs this may not be all that surprising but it is worth noting that by far the vast majority of 400- and 
500-level responses come from non-OA DOIs. Given our dataset, this observation indicates that OA content providers 
show more consistency across the board compared to non-OA providers and their positive effect to the overall picture 
(Figure \ref{fig:final_codes_by_DOI_ext}) is visible. A larger scale analysis of OA versus non-OA content providers 
is needed, however, to more reliably underline this observation. We leave such effort for future work.
\begin{table}[t!]
\centering
\caption{Distribution of DOIs leading to $OA$ and $nOA$ resources as well as to $SUB$ and $nSUB$ content in our dataset.}
\begin{tabular}{|c||c|c||c||c|c|} \hline
& \textbf{OA} & \textbf{nOA} & & \textbf{SUB} & \textbf{nSUB} \\ \hline \hline
$DOI_{ext}$ & $973$ & $9,027$ & $DOI_{int}$ & $1,266$ & $8,734$ \\ \hline
\end{tabular}
\label{tab:access_levels}
\end{table}

\begin{figure}[t!]
    \centering
    \begin{subfigure}{0.49\textwidth}
        \includegraphics[scale=0.25]{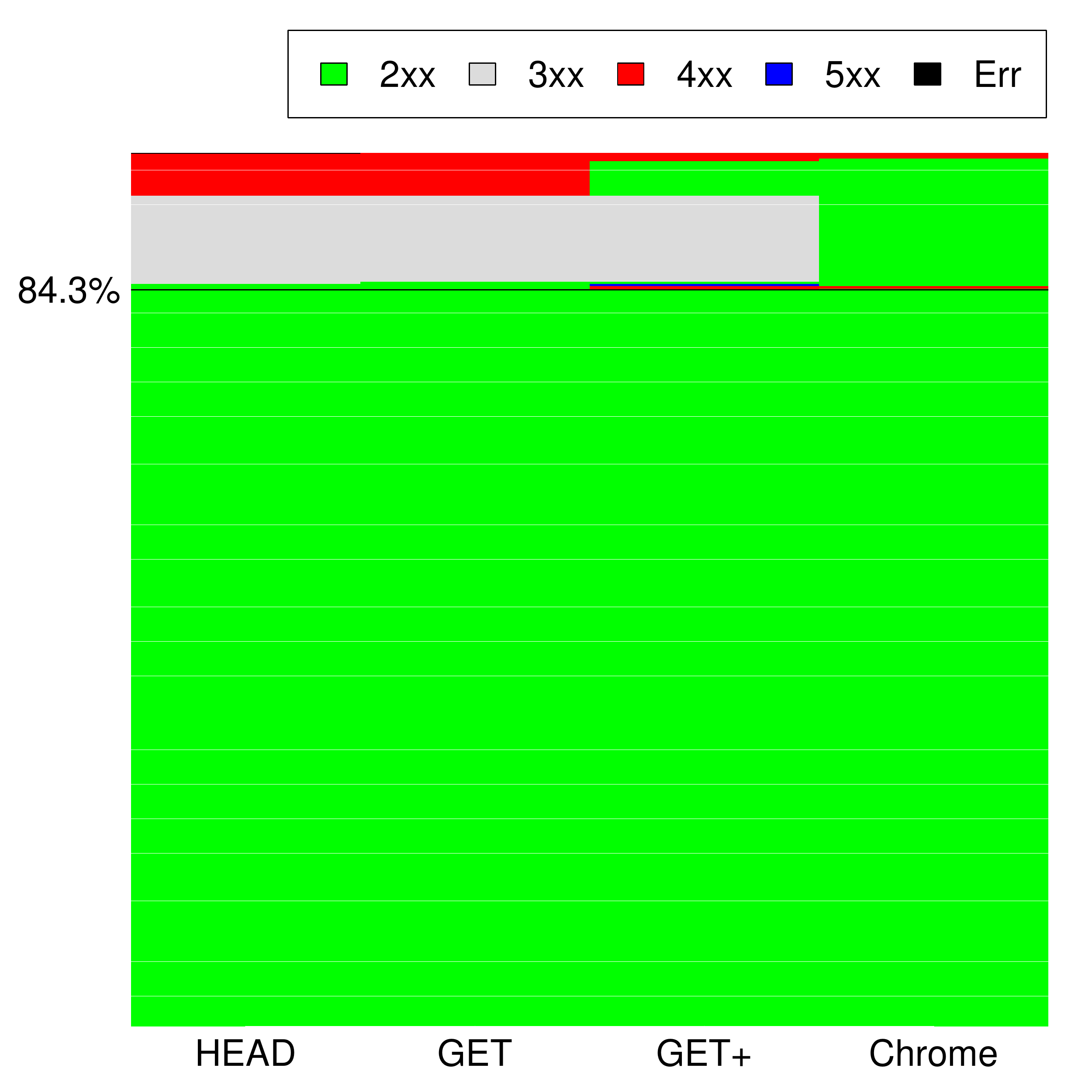}
        \caption{Subscription articles}
        \label{fig:final_codes_by_DOI_int_SUB}
    \end{subfigure}
    %~
    \begin{subfigure}{0.49\textwidth}
        \includegraphics[scale=0.25]{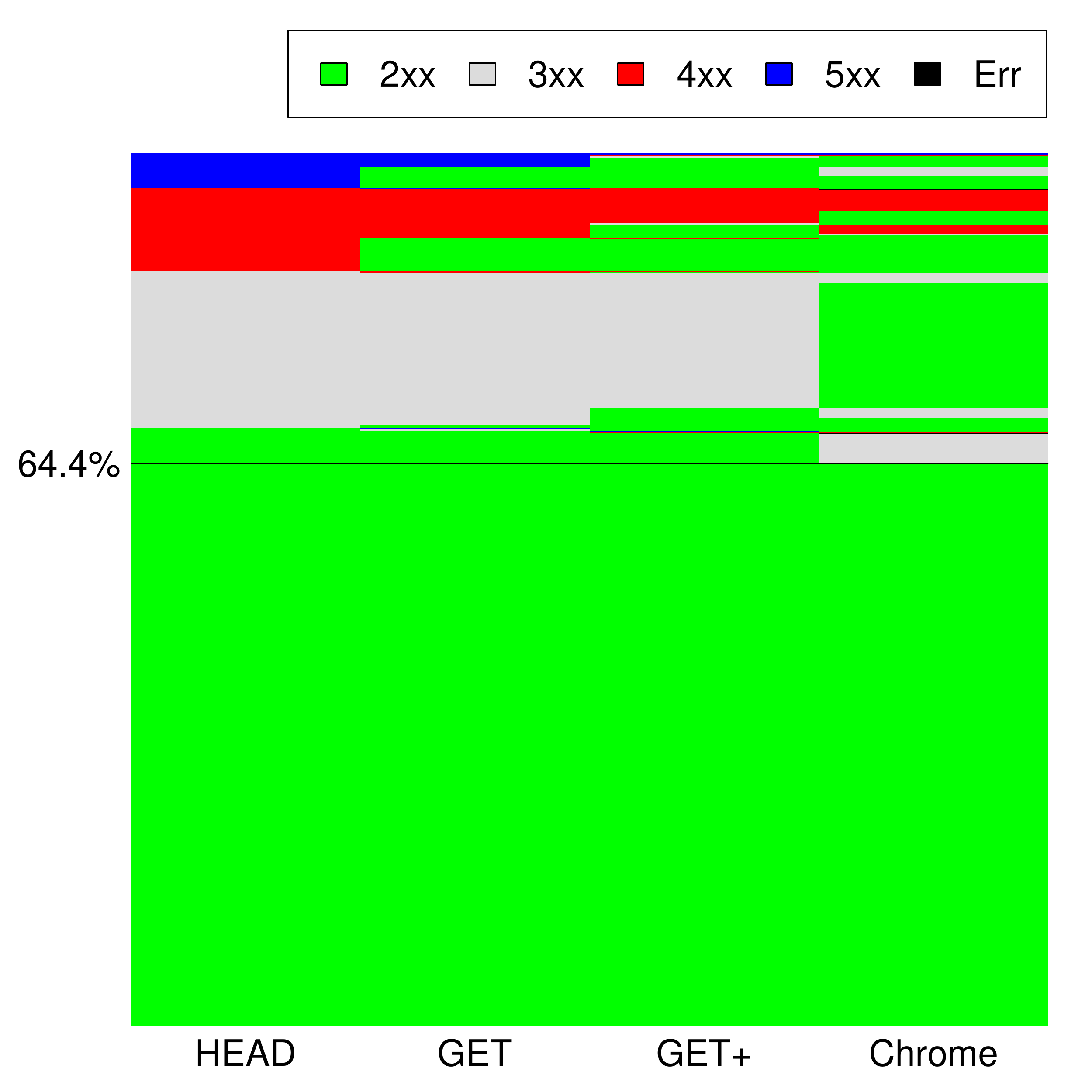}
        \caption{Non-subscription articles}
        \label{fig:final_codes_by_DOI_int_nSUB}
    \end{subfigure}
    \caption{$DOI_{int}$ final HTTP response codes distinguished by SUB and nSUB}
    \label{fig:final_codes_by_DOI_int_SUBnSUB}
\end{figure}

Figure \ref{fig:final_codes_by_DOI_int_SUBnSUB} shows the final response codes for DOIs that identify institutionally 
licensed content (Figure \ref{fig:final_codes_by_DOI_int_SUB}) and content not licensed by our institution 
(Figure \ref{fig:final_codes_by_DOI_int_nSUB}). We see a much higher ratio of DOIs returning 200-level
responses for all request methods for licensed content ($84.3\%$) compared to not licensed content ($64.4\%$). 
We also notice fewer 300-, 400-, and 500-level responses for licensed content and the Chrome method being almost 
perfect in returning 200-level responses ($99\%$).
When we again compare Figure \ref{fig:final_codes_by_DOI_int_SUBnSUB} to the overall picture for this corpus shown in 
Figure \ref{fig:final_codes_by_DOI_int}, we notice a strong resemblance between 
Figures \ref{fig:final_codes_by_DOI_int_nSUB} and \ref{fig:final_codes_by_DOI_int}. 
This leads us to conclude that providers, when serving licensed content, show more consistency and introduce fewer 
unsuccessful DOI resolutions.
\section{Conclusions} \label{sec:conclusions}
In this paper we investigate the notion of persistence of DOIs as persistent identifiers from the
perspective of their resolution on the web. 
Based on a previously generated corpus of DOIs and enhanced by an additional corpus of popular URIs, 
we present our results from dereferencing these resources with four very common but different HTTP request 
methods. We report on HTTP response codes, redirect chain length, and response code changes and highlight 
observed differences for requests originating from an external and internal network. We further analyze 
the effect of Open Access versus non-Open Access and licensed versus not licensed content.
We expected the resolution of DOIs to be consistent but our findings do not show a consistent picture at all. 
More than half of all requests ($51.7\%$) are unsuccessful from an external network compared to just over 
$33\%$ from an institutional network. In addition, the success rate varies across request methods. We find that
the method that most closely resembles the human browsing behavior (Chrome method) generally works best.
We observed an alarming amount of changes in response code depending on the HTTP request method used. These
findings provide strong indicators that scholarly content providers reply to DOI requests differently, 
depending on the request method, the originating network environment, and institutional subscription levels. 
Our scholarly record, to a large extend, relies on DOIs to persistently identify scholarly resources on the web. 
However, given our observed lack of consistency in DOI resolutions on the publishers' end, we raise serious 
concerns about the persistence of these persistent identifiers of the scholarly web. 
%
%
% ---- Bibliography ----
%

%
%
\end{document}